\documentclass{elsart}
\usepackage{natbib}
\usepackage{amsmath,amssymb,graphicx}

\newcommand{\be}{\begin{equation}}
\newcommand{\ee}{\end{equation}}
\newcommand{\bea}{\begin{eqnarray}}
\newcommand{\eea}{\end{eqnarray}}

\begin{document}

\begin{frontmatter}
\title{Coherent dynamics on hierarchical systems}
\author{Alexander Blumen, Veronika Bierbaum, and Oliver M{\"u}lken}
\address{Theoretische Polymerphysik, Universit\"at Freiburg,
Hermann-Herder-Stra{\ss}e 3, D-79104 Freiburg i.Br., Germany}

\begin{abstract}
We study the coherent transport modeled by continuous-time quantum walks,
focussing on hierarchical structures. For these we use Husimi cacti,
lattices dual to the dendrimers. We find that the transport depends
strongly on the initial site of the excitation. For systems of sizes
$N\le21$, we find that processes which start at central sites are nearly
recurrent.  Furthermore, we compare the classical limiting probability
distribution to the long time average of the quantum mechanical transition
probability which shows characteristic patterns. We succeed in finding a
good lower bound for the (space) average of the quantum mechanical
probability to be still or again at the initial site.
\end{abstract}

\end{frontmatter}

{\bf Introduction}\\
The dynamics of excitons is a problem of long standing in molecular and
polymer physics \cite{Kenkre}. Thus, the incoherent exciton transport can
be efficiently modeled by random walks, see, for instance,
\cite{heijs2004,blumen2005}; then, the transport follows a master (rate)
equation and the underlying topology is fundamental for the
dynamics \cite{weiss}.  Studying the coherent transport, we model the
process using Schr\"odinger's equation, which is mathematically closely
related to the master equation, where the transfer rates enter through the
connectivity matrix ${\bf A}$ of the structure
\cite{farhi1998,mb2005a,mvb2005a}. Moreover, also the elements of the
secular matrix in H\"uckel's theory are given by ${\bf A}$
\cite{McQuarrie}.

\begin{figure}[htb]
\centerline{
\includegraphics[clip=,width=\columnwidth]{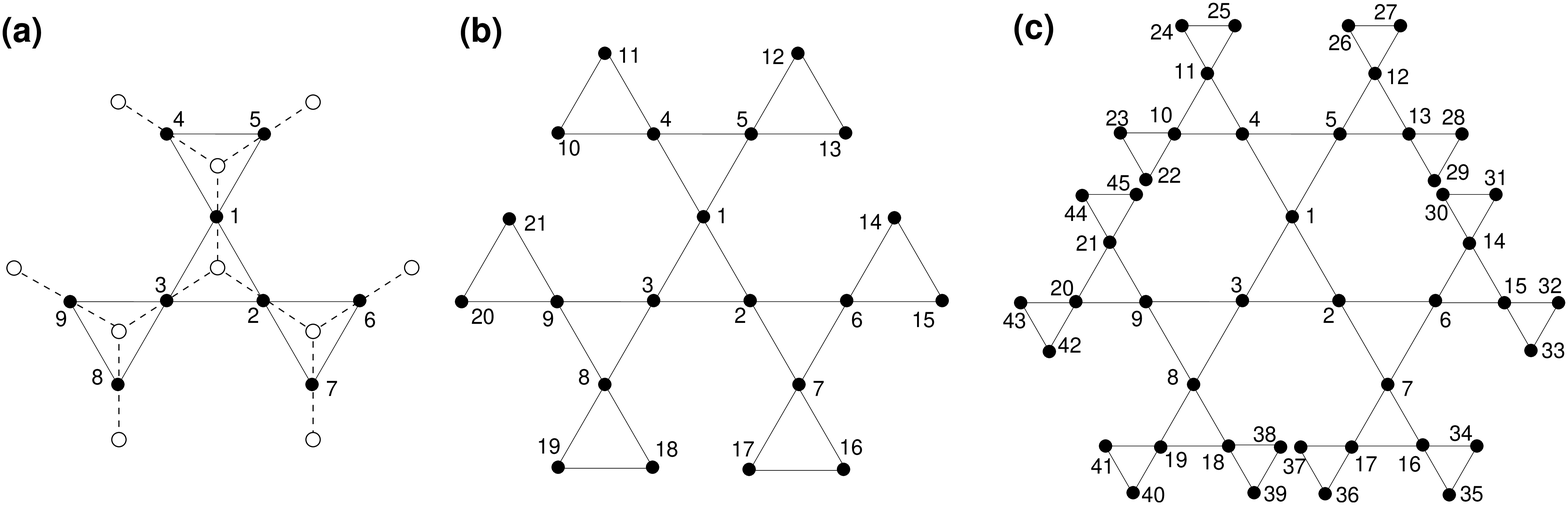}}
\caption{Finite Husimi cacti (filled circles) of size (a) $N=9$, (b)
$N=21$, and (c) $N=45$. (a) also shows with dashed lines and open circles
the corresponding dendrimer.
}
\label{husimicactus}
\end{figure}

Hyperbranched macromolecules have attracted a lot of attention in recent
years. In this respect dendrimers have served as a prime example
\cite{Voegtle}. Among a series of very interesting and crossdisciplinary
applications like drug delivery, the role of dendrimers as light
harvesting antennae has been investigated
\cite{mukamel1997,jiang1997,barhaim1997,barhaim1998}. The details of the
transport of an optical excitation over these structures depend on the
localized states used. In one picture one may envisage the excitation to
occupy the branching points of the dendrimer, see, for instance,
\cite{barhaim1997,barhaim1998,mbb2006a}.  However, as shown in
\cite{poliakov1999}, it may happen that the exciton occupies
preferentially the bonds between the branching points. Then, the essential
underlying structure is different and is given by sites localized at the
mid-points of the bonds.  We exemplify the situation in
Fig.~\ref{husimicactus}(a), starting from a dendrimer of generation $2$
(open circles) and indicating the mid-points of the bonds by filled
circles. Connecting neighboring filled circles by new bonds, one is led to
the dual lattice of the dendrimer, which is called a Husimi cactus. Figure
\ref{husimicactus} shows three finite Husimi cacti of sizes $N=9,21$, and
$45$.

{\bf Coherent dynamics modelled by quantum walks}\\
The quantum mechanical extension of a continuous-time random walk (CTRW)
on a network (graph) of connected nodes is called a continuous-time
quantum walk (CTQW). It is obtained by identifying the Hamiltonian of the
system with the (classical) transfer matrix, ${\bf H} = - {\bf T}$, see
e.g.\ \cite{farhi1998,mb2005a} (we will set $\hbar \equiv 1$ and
$m\equiv1$ in the following). The transfer matrix of the walk, ${\bf T} =
(T_{ij})$, is related to the connectivity matrix ${\bf A}$ of the graph by
${\bf T} = - \gamma {\bf A}$, where for simplicity we assume the
transmission rates $\gamma$ of all bonds to be equal and set
$\gamma\equiv1$ in the following.  The matrix ${\bf
A}$ has as non-diagonal elements $A_{ij}$ the values $-1$ if nodes $i$ and
$j$ of the graph are connected by a bond and $0$ otherwise.  The diagonal
elements $A_{ii}$ of ${\bf A}$ equal the number of bonds, $f_i$, which exit
from node $i$.

The basis vectors $|j\rangle$ associated with the nodes $j$ span the whole
accessible Hilbert space to be considered here. The time evolution of a
state $| j \rangle$ starting at time $t_0$ is given by $| j;t \rangle =
{\bf U}(t,t_0) | j \rangle$, where ${\bf U}(t,t_0) = \exp[-i {\bf H}
(t-t_0)]$ is the quantum mechanical time evolution operator.  The
transition amplitude $\alpha_{k,j}(t)$ from state $| j \rangle$ at time
$0$ to state $|k\rangle$ at time $t$ reads then $\alpha_{k,j}(t) = \langle
k | {\bf U}(t,0) | j \rangle$ and obeys Schr\"odinger's equation.
Denoting the orthonormalized eigenstates of the Hamiltonian ${\bf H} =
-{\bf T}$ by $| q_n\rangle$ (such that $\sum_n | q_n\rangle \langle  q_n |
= \boldsymbol 1$) and the corresponding eigenvalues by $\lambda_n$, the
quantum mechanical transition probability (TP) is
\be
\pi_{k,j}(t) \equiv |\alpha_{k,j}(t)|^2 = \left| \sum_n \langle k| e^{-i
\lambda_n t} | q_n\rangle
\langle  q_n | j
\rangle \right|^2.
\label{qm_prob_full}
\ee
Note that for the corresponding classical probabilities $\sum_k p_{k,j}(t)
= 1$ holds, with $p_{k,j}(t) = \sum_n \langle k| e^{- \lambda_n t} |
q_n\rangle \langle  q_n | j \rangle$, whereas quantum mechanically we have
$\sum_k |\alpha_{k,j}(t)|^2 = 1$.

{\bf Transition probabilities}\\
In the following we present the TPs $\pi_{k,j}(t)$ obtained from
diagonalizing the Hamiltonian ${\bf H}$ by using the standard software
package MATLAB.  For the finite Husimi cactus consisting of $21$ nodes, as
depicted in Fig.~\ref{husimicactus}, we find numerically that the TPs are
nearly periodic when the initial excitation is placed on one of the
(symmetrically equivalent) nodes $1$, $2$, or $3$ of the inner triangle.
Thus at time $t=35.551846$ one has $\pi_{1,1}(35.551846)= 0.998761$, when
starting from $\pi_{1,1}(0)=1$. In fact, the following expression holds
exactly for all $k$ sites of the cactus with $N=21$:
\be
\pi_{k,1}(t) = \left[\sum_{l=1}^6 a_l^{(k)} \cos(\lambda_l
t) \right]^2 + \left[\sum_{l=1}^6 a_l^{(k)} \sin(\lambda_lt)
\right]^2.
\label{alpha_recurr}
\ee
Here $\lambda_1=0$, $\lambda_2=3$, $\lambda_3=3+\sqrt{2}$,
$\lambda_4=3-\sqrt{2}$, $\lambda_5=3+2\sqrt{2}$, $\lambda_6=3-2\sqrt{2}$,
and the $a_l^{(k)}$ follow from the corresponding eigenvectors. We further
remark that the $N=21$ cactus has $8$ distinct eigenvalues, but that not
all of them enter $\pi_{k,1}(t)$.

\begin{figure}[htb]
\centerline{\includegraphics[clip=,width=\columnwidth]{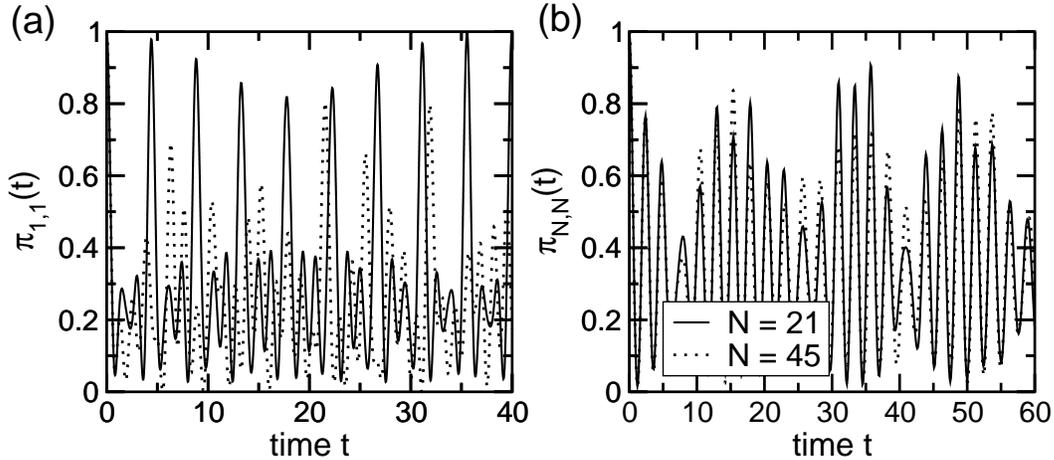}}
\caption{Transition probabilities $\pi_{j,j}(t)$ of the transport over
finite Husimi cacti of sizes $N=21$ (full lines) and $N=45$ (dashed lines)
for (a) $j=1$ and (b) $j=N$.
}
\label{husimi_innen}
\end{figure}

Figure~\ref{husimi_innen} shows the probability to be at the initial node
after time $t$ for two different sizes $N$ and two different initial
conditions. For $N=21$, when the excitation starts at the core ($j=1$),
the TP $\pi_{1,1}(t)$ is nearly recurrent. This is not anymore the case
for $N=45$.  Moreover, when the excitation starts at the periphery
($j=N$), the TPs $\pi_{N,N}(t)$ are even more irregular. Now the
$\pi_{k,N}(t)$ for $N=21$ with $j=N$ indeed also contain the two
eigenvalues $\lambda_7=3+\sqrt{6}$ and $\lambda_8=3-\sqrt{6}$ in addition
to those given after Eq.~(\ref{alpha_recurr}), a fact which increases the
irregularity of the temporal behavior of the $\pi_{k,N}(t)$.

\begin{figure}[htb]
\centerline{\includegraphics[clip=,width=\columnwidth]{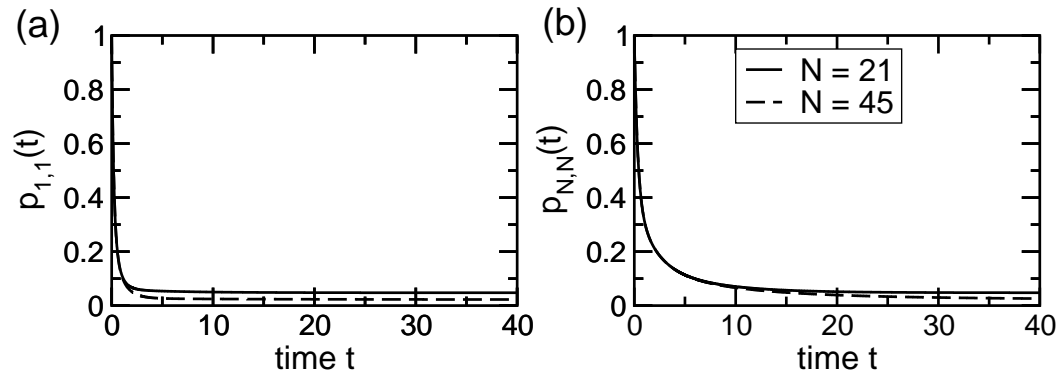}}
\caption{Classical transition probabilities $p_{j,j}(t)$ of the transport
over finite Husimi cacti of sizes $N=21$ (full lines) and $N=45$ (dashed
lines) for (a) $j=1$ and (b) $j=N$.
}
\label{husimi_innen_cl}
\end{figure}

For comparison we also show in Fig.~\ref{husimi_innen_cl} the classical
transition probabilities $p_{j,j}(t)$ for the same initial conditions as
in the previous CTQW. Also here, there is a difference between the
starting points. If the excitation starts at the central node $1$, the
equipartitioned limiting probability $1/N$ is reached much faster than
when starting at the periphery.

{\bf Long time limit}\\
The quantum mechanical time evolution is symmetric to inversion. This
prevents $\pi_{k,j}(t)$ from having a definite limit for $t\to\infty$. In
order to compare the classical long time probability with the quantum
mechanical one, one usually uses the limiting probability (LP)
distribution \cite{aharonov2001} 
\be
\chi_{k,j} \equiv \lim_{T\to\infty} \frac{1}{T} \int_0^T dt \
\pi_{k,j}(t),
\label{limprob}
\ee
which can be rewritten by using the orthonormalized eigenstates of the
Hamiltonian, $| q_n\rangle$, as \cite{mvb2005a,mbb2006a}
\be
\chi_{k,j} = 
\sum_{n,m}
\delta_{\lambda_n,\lambda_m}
\langle k |
q_n \rangle \langle q_n |
j \rangle \langle j | q_m
\rangle \langle q_m | k \rangle.
\label{limprob_ev}
\ee
Some eigenvalues of $\bf H$ might be degenerate, so that the sum in
Eq.~(\ref{limprob_ev}) can contain terms belonging to different
eigenstates $| q_n\rangle$ and $| q_m\rangle$.

\begin{figure}[htb]
\centerline{
\includegraphics[clip=,width=0.5\columnwidth]{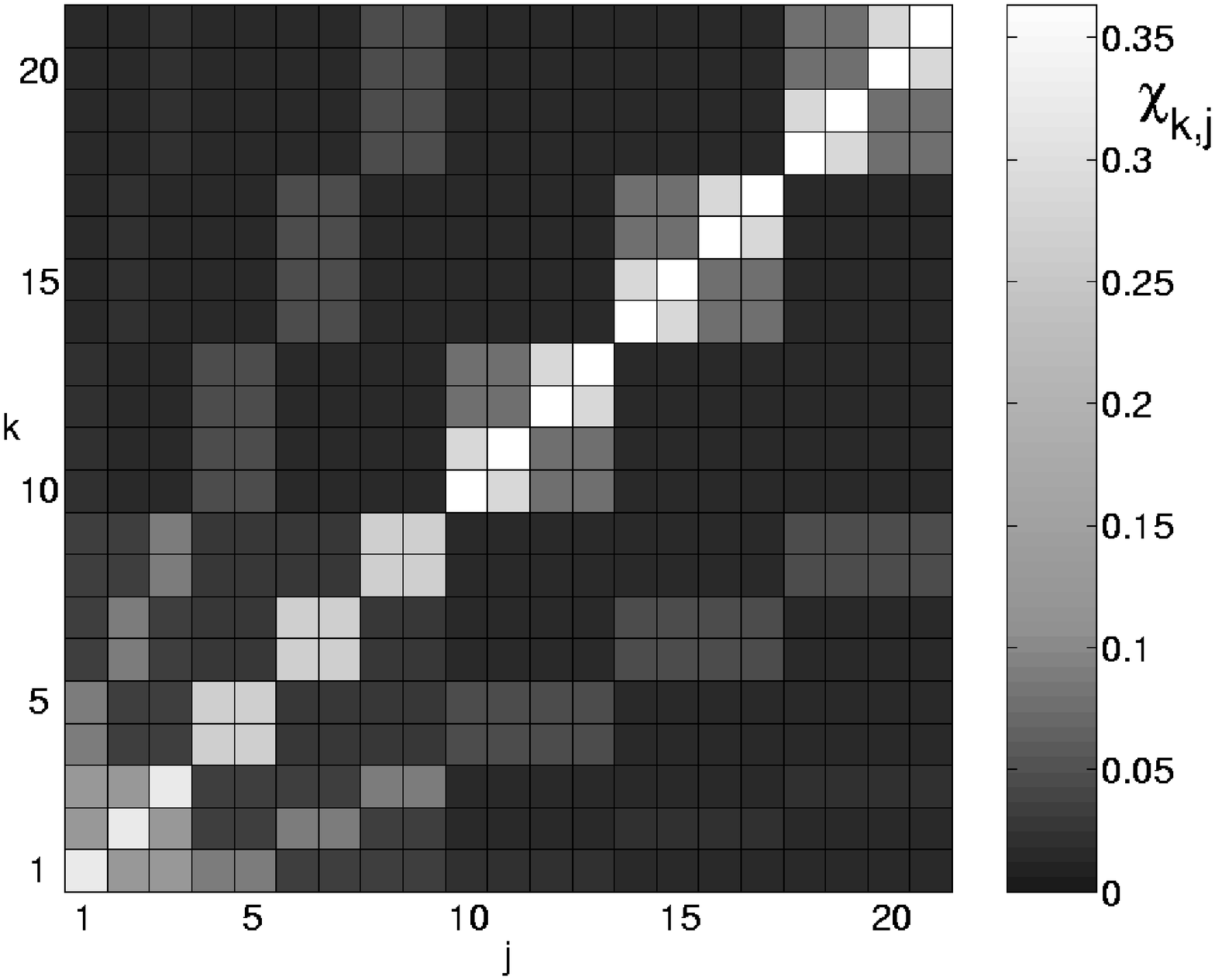}
\includegraphics[clip=,width=0.5\columnwidth]{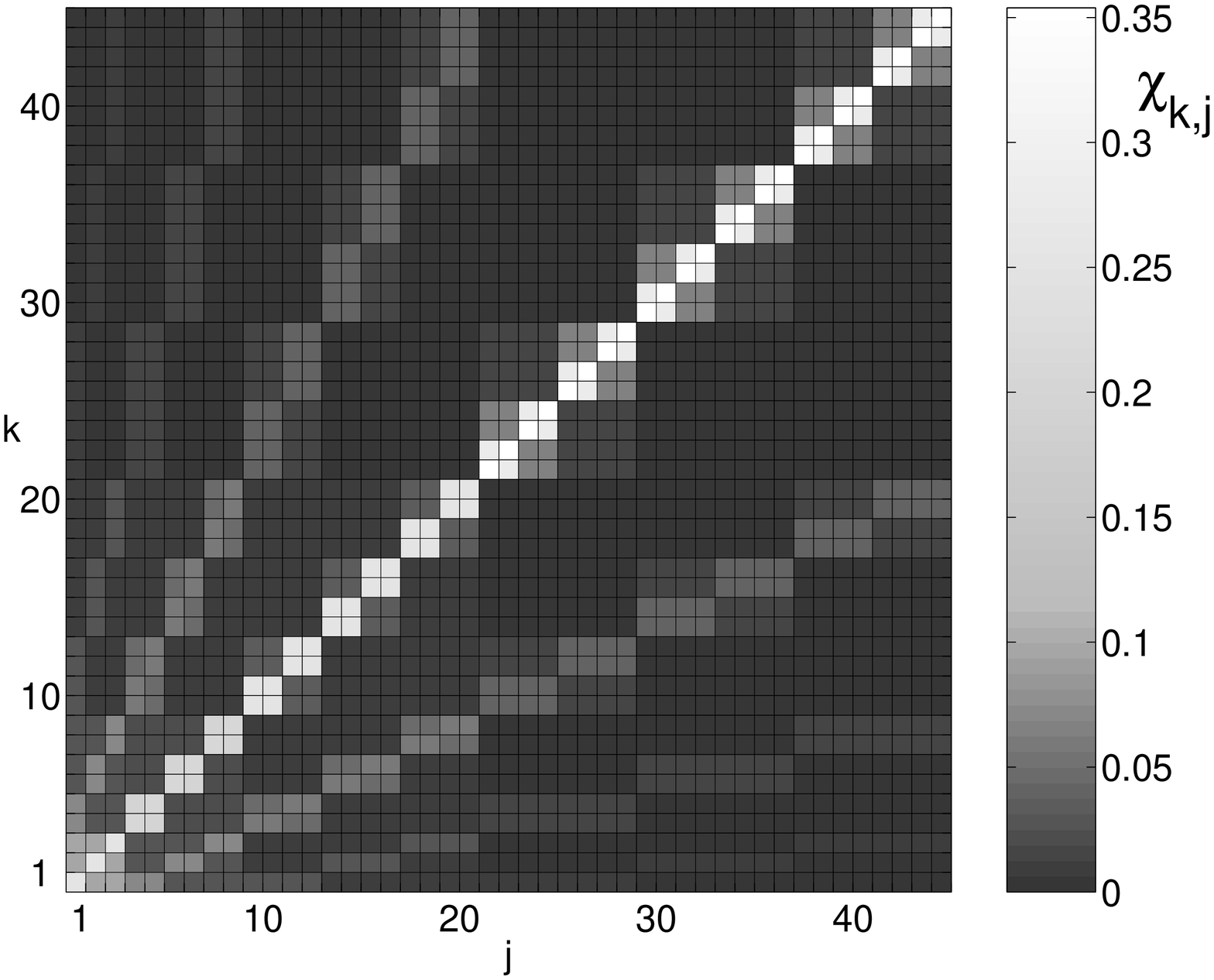}}
\caption{Limiting probabilites for finite Husimi cacti of sizes $N=21$
(left) and $N=45$ (right).}
\label{limiting}
\end{figure}

In Fig.~\ref{limiting} we present the LPs for two sizes of Husimi cacti,
$N=21$ and $N=45$. The LP distributions are self-similar generation after
generation. Furthermore, for each size there are LPs having the same
value, i.e.\ $\chi_{k,j} = \chi_{l,j}$. We collect LPs of the same value
into clusters. Depending on where the excitation starts, the clustering is
different. For instance, for $N=21$, with an excitation starting at node
$1$, there are $6$ different clusters, namely node $1$ forms one cluster
and the nodes $\{2,3\}$, $\{4,5\}$, $\{6,7,8,9\}$, $\{10,11,12,13\}$, and
$\{14,\cdots,21\}$ the remaining $5$ clusters. When the excitation starts
at another node, the clusters are formed by different nodes, as can be
infered from Fig.~\ref{limiting}. 

These results are closely related to our findings for the coherent
transport over dendrimers \cite{mbb2006a}. For dendrimers the grouping of
sites into clusters also depends on the starting site.  Furthermore, there
exists a lower bound for the LPs; one has namely $\chi_{k,j}\ge1/N^2$ for
all nodes $k$ and $j$ \cite{mbb2006a}. This lower bound is related to the
fact that in our cases ${\bf A}$ has exactly one vanishing eigenvalue,
$\lambda_0=0$.    

{\bf Averaged probabilities}\\
As discussed above, starting at a central node of the $N=21$ cactus, say
node $1$, leads to nearly periodic $\pi_{k,1}(t)$.  However, for larger
cacti and/or different initial conditions this does not have to be the
case.

Classically one has a very simple expression for the probability to be
still or again at the initially excited node, {\sl spacially} averaged
over all nodes. Then one finds \cite{blumen2005}
\be
\overline{p}(t) \equiv \frac{1}{N} \sum_{j=1}^{N} p_{j,j}(t) =
\frac{1}{N} \sum_{n=1}^{N} \ \exp\big(-\lambda_n t\big).
\ee
This result is quite remarkable, since it depends only on the eigenvalue
spectrum of the connectivity matrix, but {\sl not} on the eigenvectors.

Quantum mechanically, we obtain a lower bound by using the Cauchy-Schwarz
inequality, i.e., \cite{mbb2006a},
\be
\overline{\pi}(t) \equiv \frac{1}{N} \sum_{j=1}^{N} \pi_{j,j}(t)\geq
\frac{1}{N^2} \sum_{n,m} \ \exp\big[-i(\lambda_n - \lambda_m)t\big].
\label{csi}
\ee
In analogy to the classical case, the lower bound in Eq.~(\ref{csi})
depends only on the eigenvalues and {\sl not} on the eigenvectors. Note
that for a CTQW on a simple regular network with periodic boundary
conditions the lower bound is exact.  We sketch the proof for a $1$D
regular network of length $N$. There, the transition amplitudes to be
still or again at node $j$ after time $t$ read $\alpha_{j,j}(t) = N^{-1}
\sum_{n} \exp(-i \lambda_{n} t)$, see Eq.~(16) of \cite{mb2005b} with
$\lambda_n = 2- 2\cos(2\pi n/N)$. The average $\overline{\pi}(t)$ is then
given by
\be
\overline{\pi}(t) = \frac{1}{N} \sum_{j=1}^N \left| \frac{1}{N}
\sum_{n} \exp(-i \lambda_{n} t) \right|^2 = \frac{1}{N^2}
\sum_{n,m} \ \exp\big[-i(\lambda_n - \lambda_m)t\big].
\label{bound_pbc}
\ee
This result also holds for hypercubic lattices in higher, $d$-dimensional
spaces, when the problem separates in every direction; then one has
$\alpha^{(d)}_{j,j}(t) = [\alpha_{j,j}(t)]^d$ (for the $d=2$ case see
\cite{mvb2005a}). 

\begin{figure}[htb]
\centerline{\includegraphics[clip=,width=\columnwidth]{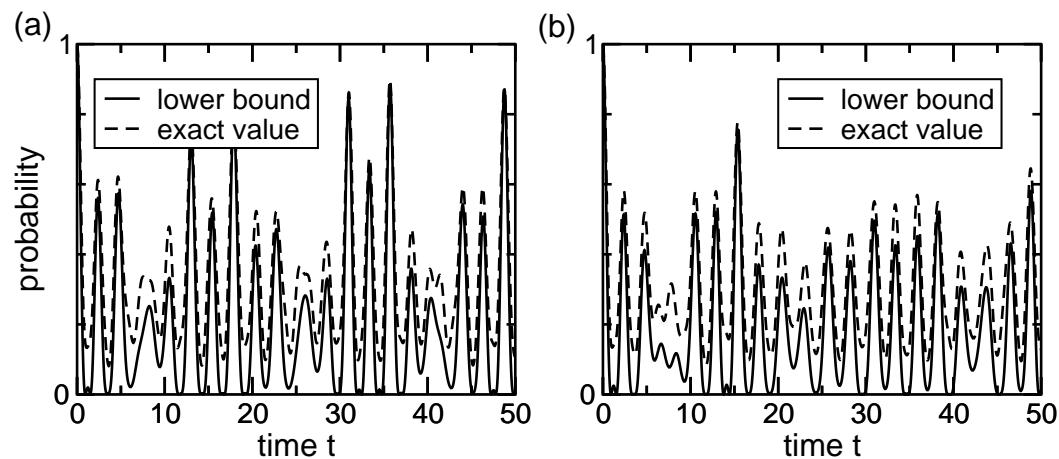}}
\caption{Comparison of the averaged probability to be still or again at
the initial node with the lower bound given in Eq.~(\ref{csi}) for finite
Husimi cacti of sizes (a) $N=21$ and (b) $N=45$.}
\label{husimicsu}
\end{figure}

In Fig.~\ref{husimicsu} we compare the exact value of $\overline{\pi}(t)$
with the lower bound given in Eq.~(\ref{csi}).  We use cacti of two
different sizes, namely $N=21$ and $N=45$. Now, the lower bound fluctuates
more than the exact curve, but it reproduces quite well the overall
behavior of the extrema of the exact curve. Especially the strong maxima
of the exact value are quantitatively reproduced.  This is in accordance
with previous studies on exciton transport over square lattices
\cite{mvb2005a} and over dendrimers \cite{mbb2006a}. 

{\bf Conclusions}\\
In conclusion, we have modeled the coherent dynamics on finite Husimi
cacti by continuous-time quantum walks. The transport is only determined
by the topology, i.e.\ by the connectivity matrix, of the cacti. For the
$N=21$ cactus the dynamics is nearly periodic when one of the central nodes
is initially excited. For larger structures and/or different initial
conditions we observe only partial recurrences.  To compare these results
to those of the classical (incoherent) case, we calculated the long time
average of the TPs.  Depending on the initial
conditions, these show characteristic patterns, by which different nodes
displaying the same limiting probabilities may be grouped into clusters.
Furthermore, we calculated a lower bound for the average probability to be
still or again at the initial node after some time $t$. This lower bound
depends only on the eigenvalues of ${\bf A}$ and agrees quite well with
the exact value.

{\bf Acknowledgments}\\
This work was supported by a grant from the Ministry of Science, Research
and the Arts of Baden-W\"urttemberg (Grant No.\ AZ: 24-7532.23-11-11/1).
Further support from the Deutsche Forschungsgemeinschaft (DFG) and the
Fonds der Chemischen Industrie is gratefully acknowledged.

\end{document}